\documentclass{PoS}

\def\3{\ss{}}

\def\d{{\rm d}}

\newcommand{\beq}{\begin{equation}}
\newcommand{\eeq}{\end{equation}}
\newcommand{\bea}{\begin{eqnarray}}
\newcommand{\eea}{\end{eqnarray}}

\title{Direct detection of neutralino dark matter with DM@NLO}

\ShortTitle{Direct detection of neutralino dark matter with DM@NLO}

\author{\speaker{Michael Klasen}
 \thanks{Supported by BMBF under contract 05H15PMCCA.}\\
 Institut f\"ur Theoretische Physik, Westf\"alische
 Wilhelms-Universit\"at M\"unster, Wilhelm-Klemm-Stra\ss{}e 9,
 D-48149 M\"unster, Germany\\
 E-mail: \email{michael.klasen@uni-muenster.de}}

\author{Karol Kova\v{r}\'ik%
 \\
 Institut f\"ur Theoretische Physik, Westf\"alische
 Wilhelms-Universit\"at M\"unster, Wilhelm-Klemm-Stra\ss{}e 9,
 D-48149 M\"unster, Germany\\
 E-mail: \email{karol.kovarik@uni-muenster.de}}

\author{Saskia Schmiemann%
 \thanks{Supported by DFG under contract GRK 2149.}\\
 Institut f\"ur Theoretische Physik, Westf\"alische
 Wilhelms-Universit\"at M\"unster, Wilhelm-Klemm-Stra\ss{}e 9,
 D-48149 M\"unster, Germany\\
 E-mail: \email{saskia.schmiemann@uni-muenster.de}}

\abstract{
\vspace*{-160mm}
\flushright{\large MS-TP-17-13}\\
\vspace*{150mm}
 We calculate spin-independent and spin-dependent direct detection
 cross sections of neutralino dark matter at next-to-leading order of QCD.
 The numerical effects are comparable in size to the uncertainties in the
 nuclear matrix elements. Our results are applicable to bino, wino or higgsino
 dark matter and allow for consistent correlations with the relic density in
 DM@NLO.}

\FullConference{EPS-HEP 2017, European Physical Society conference on High Energy Physics\\
		5-12 July 2017\\
		Venice, Italy}

\begin{document}

\section{Direct detection of neutralino dark matter}

The identification of dark matter (DM) is one of the most urgent questions
in astroparticle physics. For many decades, evidence for its sizeable
presence in the Universe and its important role in structure formation has
been accumulating, and the overall relic density $\Omega h^2$ ($h$ being the
present Hubble expansion rate in units of 100 km s$^{-1}$ Mpc$^{-1}$) has
been precisely measured \cite{Ade:2015xua}. The
lightest neutral supersymmetric (SUSY) partner of electroweak gauge
and Higgs bosons (neutralino $\tilde{\chi}^0_1$) continues to be a prime
candidate for WIMP (weakly interacting massive particle) DM, and
theoretical calculations of the DM relic density at next-to-leading
order (NLO) of QCD with DM@NLO, which include all coannihilation channels
(with the exception of $\tilde{t}_1\tilde{t}_1\to q\bar{q},gg$)
now match the experimental precision \cite{Harz:2016dql}. For an unambiguous
identification of DM, it must, however, be detected on Earth, e.g.\ with
large kryogenic detectors like XENON1T \cite{Aprile:2017iyp}. Comparisons
with theoretical cross section calculations and correlations with the relic
density or other observables (e.g.\ from indirect detection or the LHC) should
then allow for a precise extraction of the DM mass and couplings. For
neutralinos, this is now possible thanks to the calculation of NLO SUSY-QCD
corrections to the neutralino-nucleon cross section and the inclusion of this
second DM observable in DM@NLO \cite{Klasen:2016qyz}.

\section{Neutralino-nucleon cross section}

The differential rate for direct DM detection (in counts/kg/day/keV)
\beq
 \frac{\mathrm{d}R}{\mathrm{d}E} = \sum_i c_i \frac{\sigma_i}{2m_{\tilde{\chi}^0_1}\mu_i^2}\rho_0 \eta_i
\eeq
is usually expressed in terms of the nuclear mass fractions $c_i$, reduced
masses $\mu_i$, local DM density $\rho_0=0.3$ GeV/cm$^3$, and velocity
integrals $\eta_i= \int_{v_{\min,i}}^{v_{\rm esc}}\d^3v \, f(\vec{v})/v$ with
$v_{\min,i}=\sqrt{m_iE/(2\mu_i^2)}$.

Since the spin-{\em independent} cross sections for each isotope in the target
\beq
 \sigma_i^{\mathrm{SI}} = \frac{\mu_i^2}{\pi}\left|Z_i g_p^{\mathrm{SI}} +(A_i-Z_i)g_n^{\mathrm{SI}}\right|^2|F_i^{\mathrm{SI}}(Q_i)|^2
\eeq
depend on the nuclear charges $Z_i$, masses $A_i$ and structure functions
$F_i^{\rm SI}$, they are often replaced by the one for a single nucleon
(assuming $g_p=g_n$) to enable a direct comparison of different experiments.
We use, however, the exact expressions
\beq
 g_N^{\mathrm{SI}} = \sum_{q} \langle N |\bar{q}q| N\rangle \alpha_{q}^{\mathrm{SI}}
\eeq
for the spin-independent four-fermion couplings. The Wilson coefficients
$\alpha_{q}^{\mathrm{SI}}$ contain the wanted information on the electroweak
interaction of DM and quarks, while the nuclear matrix elements
$ \langle N |m_q \bar{q}q| N\rangle = f_{Tq}^N m_N$ are known to be subject
to considerable uncertainties from the non-perturbative regime of QCD
\cite{Gondolo:2004sc,Belanger:2007zz,Crivellin:2013ipa}. Beyond
the tree-level, the Wilson coefficients $\alpha_q^{\mathrm{SI}}$ are, however,
also affected by (perturbative) QCD uncertainties and become related to the
nuclear matrix elements through renormalisation group equations.\footnote{The
role of
effective gluon interactions has been discussed in Ref.\ \cite{Drees:1992rr}.}

Similarly, the spin-{\em dependent} cross section
\beq
 \sigma_i^{\mathrm{SD}} = \frac{4\mu_i^2}{2J +1}\big(|g_p^{\mathrm{SD}}|^2S_{\mathrm{pp},i}(Q_i) + |g_n^{\mathrm{SD}}|^2S_{\mathrm{nn},i}(Q_i)
 + |g_p^{\mathrm{SD}}g_n^{\mathrm{SD}}|S_{\mathrm{pn},i}(Q_i)\big)
\eeq
depends on the spin structure functions $S_{NN,i}$ and spin-dependent
four-fermion couplings
\beq
g_N^{\mathrm{SD}} = \sum_{q= u,d,s}  (\Delta q)_N \alpha_{q}^{\mathrm{SD}}.
\eeq
Here, the nuclear spin $J$ is supposed to be carried mostly by the three light
quark flavours and to be isospin symmetric.\footnote{This need not be the case
as discussed in Refs.\ \cite{deFlorian:2014yva,Li:2015wca}.}

The tree-level diagrams for neutralino-quark scattering are shown in
\begin{figure}[ht]
\begin{center}
 \includegraphics[width=0.8\textwidth]{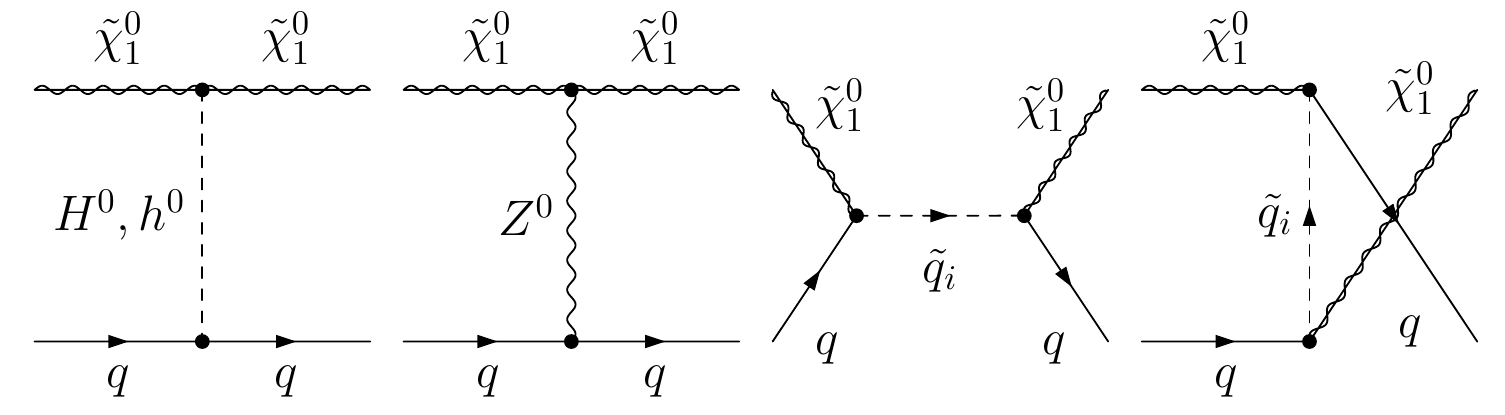}
 \caption{Full tree-level Feynman diagrams for neutralino-quark scattering.}
 \label{fig:01}
\end{center}
\end{figure}
Fig.~\ref{fig:01}. After the calculation of all self-energy, vertex and box
corrections, we renormalise the ultraviolet (UV) divergences in a mixed
on-shell and
$\overline{\rm DR}$ scheme \cite{Klasen:2016qyz}. It has the advantages of
being perturbatively stable, in particular in the top sector, and of allowing
for meaningful correlations with our relic density calculations
\cite{Harz:2016dql} and tree-level comparisons with micrOMEGAs
\cite{Belanger:2007zz}, where the same on-shell squark masses are used that
are provided by the SUSY spectrum generator SPheno \cite{Porod:2011nf}.

\begin{figure}[ht]
\begin{center}
 \includegraphics[width=0.22\textwidth]{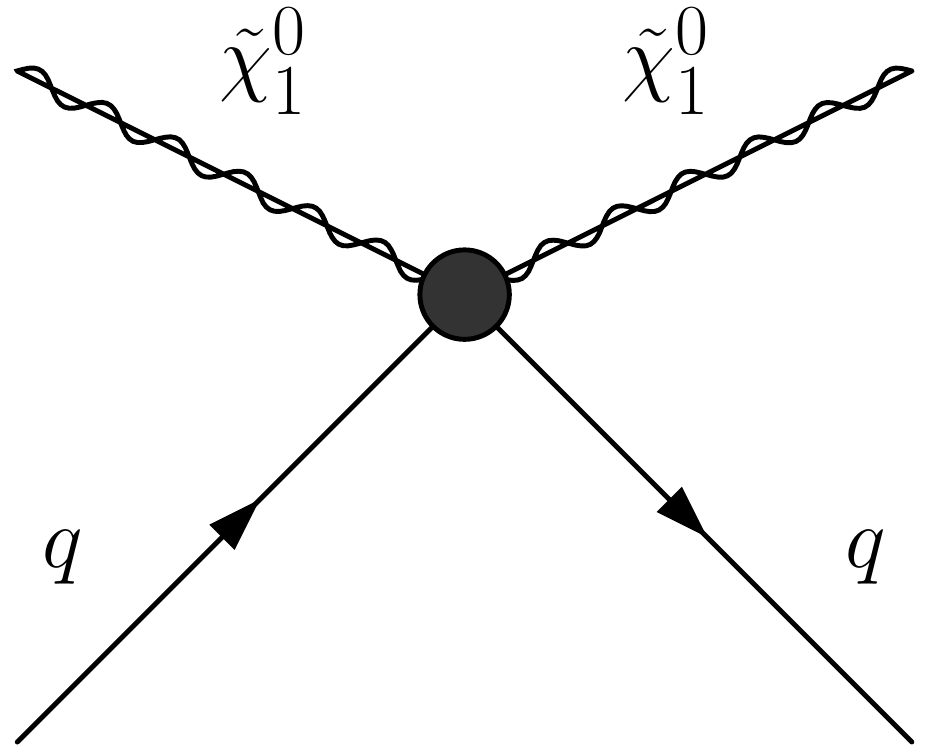}
 \caption{Effective tree-level diagram for neutralino-quark scattering.}
 \label{fig:02}
\end{center}
\end{figure}
In the non-relativistic regime, our full calculation is then matched
to the spin-independent and spin-dependent operators $Q_{1,2}$ in the
effective Lagrangian
\beq
 \mathcal{L}_\mathrm{eff} = c_1Q_1 + c_2Q_2 = c_1\bar{\chi}\chi\bar{q}q + c_2\bar{\chi}\gamma_\mu\gamma_5\chi\bar{q}\gamma^\mu\gamma_5q
\eeq
as shown symbolically in Fig.~\ref{fig:02}. As expected, the tree-level
coefficients, obtained after a Fierz transformation for the squark processes,
agree with those in DarkSUSY \cite{Gondolo:2004sc}. After the one-loop
corrections in the effective theory have also been computed, the matching
condition
\bea
 \mathcal{M}_\mathrm{full}^\mathrm{tree} + \mathcal{M}_\mathrm{full}^\mathrm{1loop} & \stackrel{!}{=} & (c_1^\mathrm{tree} + c_1^\mathrm{1loop})(Q_1^\mathrm{tree}+ Q_1^\mathrm{1loop}) + (c_2^\mathrm{tree} + c_2^\mathrm{1loop})(Q_2^\mathrm{tree} + Q_2^\mathrm{1loop})
\eea
leads to a refactorisation and UV-finite, but scale-dependent redefinitions of
Wilson coefficients and operators. In the spin-idependent case, the quark
masses $m_q(\mu)$ are factorised in $c_1$ and run from the high SUSY-breaking
scale 1 TeV to the low scale 5 GeV, where the nuclear matrix elements are
defined. In the spin-dependent case, the running of $c_2$ is given by
\beq
 \frac{c_2(\mu_\mathrm{low})}{c_2(\mu_\mathrm{high})} = \exp\left(\frac{2n_f(\alpha_s(\mu_\mathrm{high}) - \alpha_s(\mu_\mathrm{low}))}{\beta_0\pi}\right).
\eeq

\section{Numerical results}

\begin{figure}[ht]
\begin{center}
 \includegraphics[width=0.66\textwidth]{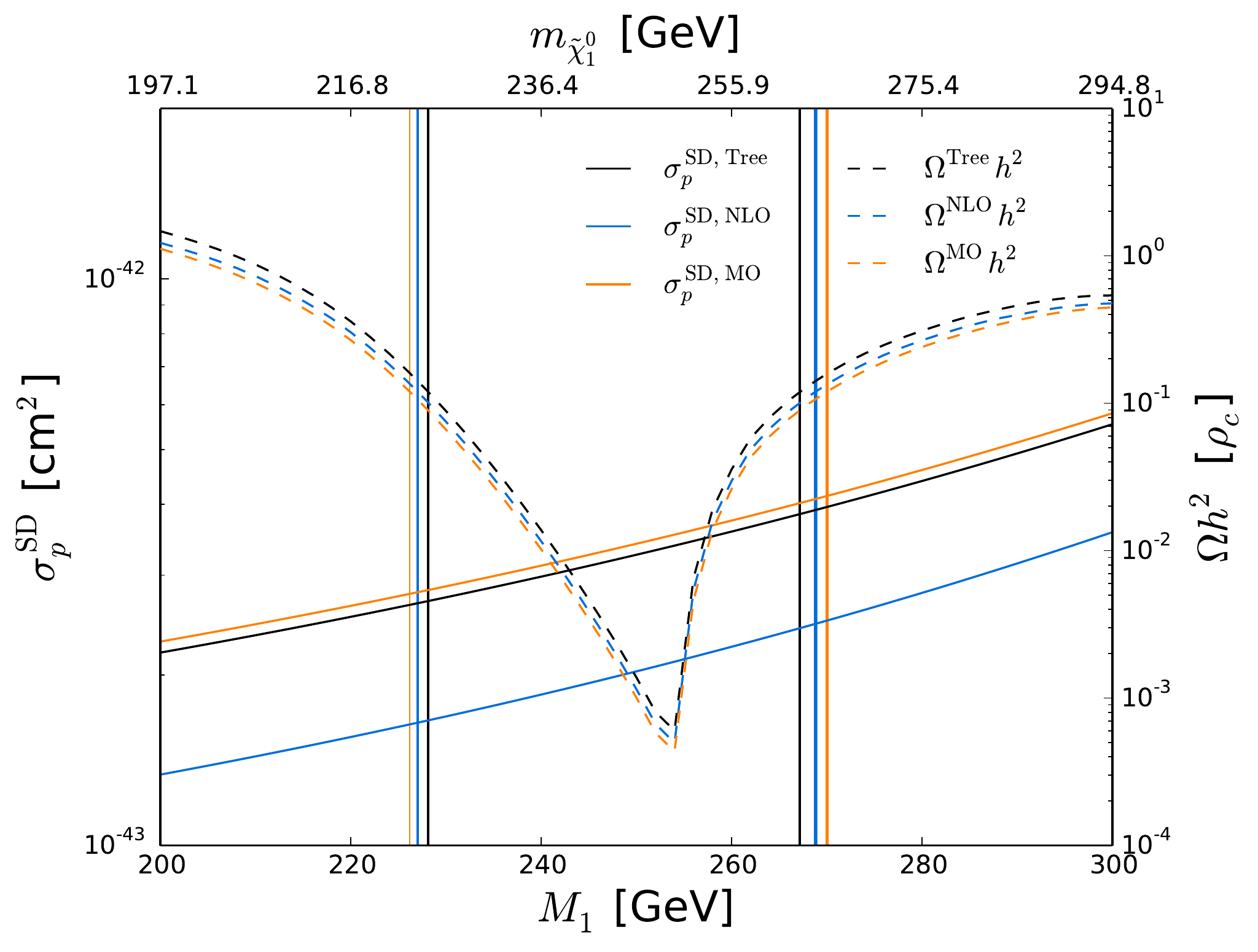}
 \caption{Combined relic density and direct detection calculation in scnenario B.}
 \label{fig:03}
\end{center}
\end{figure}

Phenomenological minimal SUSY Standard Model (pMSSM) scenarios with eleven
free parameters and bino-wino, bino-higgsino, or higgsino-bino DM, that satisfy
all current experimental constraints, have been presented in Ref.\
\cite{Harz:2016dql}. Scenario B, e.g., contains a bino-higgsino DM candidate
of about 267 GeV mass and up- and down-type squarks of mass 550 and 556 GeV,
respectively. Fig.\ \ref{fig:03} shows a scan in the bino mass parameter $M_1$
around this point, indicated there by full vertical lines at tree-level
(black), with micrOMEGAs (orange), and in NLO (blue). In the shown mass
region, a second viable scenario with a lower DM mass of about 228 GeV is
found, indicated by a second set of vertical lines. One observes good
agreement at leading order, but a significant shift at NLO in the
spin-dependent direct detection cross section (left ordinate and full curves).
The corresponding numbers are listed in Tab.\ \ref{tab:01}. When correlated
with the relic density calculations at the same order (dashed curves), this
leads to shifts in the extracted bino mass parameter $M_1$ of several GeV.
In other SUSY scenarios, these effects can even be considerably larger.
\begin{table}
 \caption{Resulting $M_1$ and spin-dependent neutralino-proton cross section when combining direct detection and relic density routines in scenario B.}
 \centering\vspace*{2mm}
 \begin{tabular}{|c|ccc|}
		\hline
			$\quad$ & $M_1$ [GeV] & $\sigma^{\mathrm{SD}}_p$ [$10^{-43}$cm$^2$]& Shift of $\sigma^{\mathrm{SD}}_p$\\ 
			\hline 
			micrOMEGAs & 226 & $2.78$ & $+3\%$ \\
			Tree level & 228 & $2.70$ &  \\	
			Full NLO &  227 & $1.65$ & $-39\%$ \\	
			\hline
			micrOMEGAs & 270 & $4.14$ & $+8\%$\\
			Tree level & 267 & $3.84$ &  \\	
			Full NLO & 269 & $2.47$ & $-36\%$ \\
			\hline
	\end{tabular}
	\label{tab:01}
\end{table}

\section{Conclusion}

In conclusion, we have summarised our recent analytical calculation of
NLO SUSY-QCD corrections to spin-independent and spin-dependent
neutralino-nucleon cross sections, emphasising our choice of renormalisation
scheme, the matching of the full diagrammatic calculation to the effective
scalar and axial-vector operators, and the renormalisation group running of
the Wilson coefficients. More technical issues like our specific tensor
reduction method, that avoids vanishing Gram determinants at non-relativistic
velocities, were omitted from our discussion, but can be found in Ref.\
\cite{Klasen:2016qyz}.

Numerical results for the direct detection of SUSY DM can now be obtained
with DM@NLO for any neutralino decomposition (bino, wino, or higgsino).
For a specific bino-higgsino benchmark scenario we found sizeable NLO
corrections, which are in fact comparable to the nuclear uncertainties,
and we demonstrated that correlations of the relic density and direct
detection rates at NLO lead to more precise determinations of the underlying
SUSY model parameters.

\end{document}